\documentclass[letterpaper,12pt]{article}

\usepackage{dgjournal}          

\usepackage{mathptmx}

\usepackage{graphics}

\usepackage[authoryear,comma,longnamesfirst,sectionbib]{natbib} 

\usepackage{amssymb}
\usepackage{amsmath}
\usepackage{amsthm}
\usepackage{mathtools}
\usepackage{setspace}
\usepackage{multirow}
\usepackage{color}
\usepackage{colortbl}
\usepackage{verbatim}
\usepackage{url}
\usepackage{ulem}

\newtheorem{prop}{Proposition}

\begin{document}

\begin{center}
\textbf{An iterative algorithm for joint covariate and random effect selection in mixed effects models}\\
Maud Delattre$^{(1)}$ and Marie-Anne Poursat$^{(2)}$
\end{center}

\begin{center}
\footnotesize{(1) UMR MIA-Paris, AgroParisTech, INRAE, Universit\'e Paris-Saclay, 75005, Paris, France.}\\
\footnotesize{(2) Universit\'e Paris-Saclay, CNRS, INRIA, Laboratoire de math\'ematiques d'Orsay, 91405, Orsay, France.}
\end{center}


\begin{abstract}
We consider joint selection of fixed and random effects in general mixed-effects models. The interpretation of estimated mixed-effects models is challenging since changing the structure of one set of effects can lead to different choices of important covariates in the model. We propose a stepwise selection algorithm to perform simultaneous selection of the fixed and random effects. It is based on BIC-type criteria whose penalties are adapted to mixed-effects models. The proposed procedure performs model selection in both linear and nonlinear models. It should be used in the low-dimension setting where the number of covariates and the number of random effects are moderate with respect to the total number of observations. The performance of the algorithm is assessed via a simulation study, that includes also a comparative study with alternatives when available in the literature. The use of the method is illustrated in the clinical study of an antibiotic agent kinetics.

\end{abstract}

\section{Introduction}

Nonlinear mixed effects models are widely used in studies where repeated measurements are observed from several independent individuals, which is common in many areas such as pharmacology, medicine or agriculture. 
Nonlinear mixed effects models can be seen as extensions of standard nonlinear regression models 
where the parameters of the model are functions of fixed and random effects and possibly individual covariates.
We adopt the hierarchical formalism of \citet{Pinheiro2009} or \citet{Lavielle2014} that includes a very broad class of mixed-effects models in which 
the $j$th observation of the $i$th subject is modeled as
\begin{equation}
y_{ij} = f(\psi_i,x_{ij}) +\varepsilon_{ij} \, , \, i=1,\ldots,N \, , \, j=1,\ldots,n_i \, ,
\label{eq:nonlinear}
\end{equation}
where the regression function $f$ can be nonlinear in at least one component of the individual parameter $\psi_i$. $N$ is the number of subjects, $n_i$ is the number of observations on subject $i$ and the $\varepsilon_{ij}$'s are residual errors. The $x_{ij}$'s are structural design variables such as time or drug dose in a pharmacological study for instance. In a mixed-effects model, the same regression function $f$ is used for the $N$ individual series of observations, but each individual has his own parameter vector $\psi_i \in \mathbb{R}^d$. The second equation of the model adds a component that describes
variability in the individual parameters within the population of $N$ subjects. The $\psi_i$'s are defined as independent Gaussian random variables such that
\begin{equation}
g(\psi_i) = \mu + \beta C_i + \eta_i \; , \; \eta_i \underset{i.i.d}{\sim} \mathcal{N}(0,\Omega),
\label{eq:phiGauss}
\end{equation}
where $C_i \in \mathbb{R}^p$ is the vector of covariates for individual $i$, $\mu$ is the intercept, $\beta$ is the $d\times p$ fixed effects matrix and $g$ is a link function. $\Omega=\left(\omega_{kl}\right)_{1\leq k,l \leq d}$ is the $d \times d$ random effects covariance matrix and may not be diagonal to allow correlations between the random effects $\eta_i$. Many examples can be found in \cite{Lavielle2014} or \cite{Pinheiro2009}.

A key issue in mixed effects modelling is to describe the different sources of variability accurately. In the model design, it is crucial to identify the relevant covariates and the relevant random effects that explain differences between individuals. In this paper, we reconsider the model design of the analysis of the kinetics of an antibiotic agent displayed in \cite{Burdet2015}. One of the objectives of this work was to characterize the pharmacokinetic variability between patients by linking the measured covariates (age, weight, gender, etc.) to the model parameters $\psi$ through the appropriate equation~\eqref{eq:phiGauss}. 
From a modelling point of view, the question is to determine if the inter-individual variability is purely random or if it is associated with known individual covariates $C_i$. Thus, the selection of important fixed or random components in the model is a fundamental problem in the analysis of pharmacokinetic data.

The identification of the nonzero components of $\beta$ and $\Omega$ is a complex issue since changing the structure of the random effects covariance matrix can lead to different, sometimes irrelevant, selections of the covariates for the fixed effects. Moreover, including unnecessary random effects could lead to a near singular covariance matrix that could be a problem for inference. In recent years, many contributions have tackled the question of simultaneous selection of covariates and random effects in mixed-effects models. For linear mixed models, \cite{Bondell2010} and \cite{Fan2012} proposed modifications of LASSO procedures (Least Absolute Shrinkage and Selection Operator (\cite{Tibshirani}) with $L_1$ penalties taken as a function of random effects. Within the framework of generalized linear mixed models, \cite{Pan2014}, \cite{chenfeipan} and \cite{Schelldorfer2014} proposed LASSO type algorithms. These methods heavily rely on the linear formalism of the models and are attractive when the number of covariates is much larger than the total number of observations. \cite{Hui2017} proposed a regularized penalized quasi-likelihood approach which is appropriate when the numbers of covariates and random effects are moderate compared to the sample size. Step-down algorithms have also been implemeted in the \verb!lmerTest! \verb!R! package (\cite{lmerTest}) for linear mixed-effects models. They are based on testing approaches that are much more difficult to extend to a nonlinear setting (\cite{Baey2019}). The framework of our study is a general nonlinear parametric class of mixed models where the number of covariates and the number of random effects is moderate with respect to the total number of observations. To the best of our knowledge, only \cite{bertrand2013} considered a broad class of nonlinear mixed-effects models. They compared bayesian methods with Lasso and HyperLasso penalty methods, but their work was dedicated to high dimensional settings for the covariates. In more standard finite dimensional settings, there is no consensus on the best model selection method in the nonlinear framework. Penalized likelihood criteria such as BIC (Bayesian Information Criterion) are very popular in many fields of applications but the appropriate definition of its penalty term was unclear until the work of \cite{Delattre2014}. Moreover, using BIC as a model selection tool suffers the drawback that the criterion needs to be evaluated for any candidate model, which can be intractable when the number of competing models is high. This is actually the case when one performs joint covariate and random effects selection in mixed-effects models, even for moderate numbers of covariates and moderate numbers of potential random effects. 

In this paper, we propose a procedure that works well in a broad range of mixed models and is feasible for rather low-dimensional problems. We have designed an iterative method inspired from stepwise regression strategies. The procedure relies at each step on appropriate BIC-type criteria. Our algorithm is implemented in a \verb!R! function which is available upon request.

The organization of the paper is outlined as follows. The general procedure for joint covariate and random effect selection is described in section 2. In section 3, a simulation study is carried out to assess the capacity of the proposed method to identify the appropriate covariates and random effects. The analysis of the real Amikacin data is then provided in section 4. Concluding remarks are given in section 5.

\section{Joint covariate and random effect selection procedure}

\subsection{Objective and approach}

Our goal is to identify the covariates and the random effects that best characterize the inter-individual variability in the nonlinear mixed effects model setting of equations \eqref{eq:nonlinear}-\eqref{eq:phiGauss}. 
The random variables of the model are the observations $\mathbf{y}=(y_{ij})_{1\leq i \leq N , 1 \leq j \leq n_i}$ and the individual parameters $\psi_i,\; i=1,\ldots,N$. Equation \eqref{eq:phiGauss} gives the decomposition of the variability of $g(\psi_i)$. The fixed effects component $\beta$ describes part of this variability by means of covariates $C_i$. The random component $\eta_i$ describes the remaining variability, i.e.,
variability between subjects that have the same covariate values.
Let us define the true \textit{covariate structure} as the set $S_{\beta}^{\star} = \left\{(k,k') \in \{1,\ldots, d\}\times \{1,\ldots, p\}:\beta_{k,k'}^{\star} \neq 0\right\}$ which contains the indices of the true nonzero fixed effects $\beta^{\star}$. As well, denote the true \textit{covariance structure}, {\it ie} the set of positions of nonzero components in the true covariance matrix $\Omega^{\star}$ of the individual parameters, by $S_{\Omega}^{\star} = \left\{(k,k')\in \{1,\ldots, d\}\times \{k,\ldots, d\}:\omega_{k,k'}^{\star} \neq 0\right\}$. Then our goal is to find $S_{\beta}^{\star}$ and $S_{\Omega}^{\star}$ simultaneously from the sampled data. When the model includes $d$ individual parameters and $p$ covariates that are measured for each individual, there are $2^{p \times d}$ possible covariate structures $\beta C_i$ to which must be added $2^d$ possible random effects structures. The number of competing models grows rapidly as $p$ and $d$ increase. If $p=6$ and $d=3$, then there are over two millions possibilities. Hence an exhaustive search becomes computationally intractable. Instead, we propose an iterative stepwise algorithm.

\subsection{Model comparison criteria}

In this section, we specify the BIC criteria that are optimized in each step of the procedure. As shown in \cite{Delattre2014}, one must be carefull with the definition of BIC in mixed effects models since the parameters may not be penalized in the same way.  The BIC penalty should be adapted to the selection objective. In order to select the optimal model, we must choose among a collection of candidate models $M$ defined by their covariate structure $S_{\beta} = \left\{(k,k') \in \{1,\ldots, d\}\times \{1,\ldots, p\}:\beta_{k,k'} \neq 0\right\}$ and by their covariance structure $S_{\Omega} = \left\{(k,k')\in \{1,\ldots, d\}\times \{k,\ldots, d\}:\omega_{k,k'} \neq 0\right\}$. Selecting the covariance structure $S_{\Omega}$ and selecting the covariate structure $S_{\beta}$ require two different BICs.

\subsubsection{Adequate partition of the model parameters}
\label{sec:partition}

The result of \cite{Delattre2014} relies on a specific partition of the parameters. The idea is to use a degenerate model formulation induced by reordering the individual parameters according to the covariance structure $S_{\Omega}$ of the model. $\psi_i$ is splitted into $\psi_{F,i}$, its non random components, and $\psi_{R,i}$, its random components. $\psi_{R,i}$ corresponds to diagonal terms $(k,k)$, $k=1,\ldots,d$ of $\Omega$ in $S_{\Omega}$, whereas $\psi_{F,i}$ corresponds to zero diagonal terms of $\Omega$. If $\psi_{F,i}$ is not empty, $\Omega$ is singular and can be written as $\Omega = \mathrm{diag}(0,\Omega_R)$. This decomposition of the individual parameters induces a particular decomposition of the model parameters :
\begin{equation}
\mu = (\mu_F,\mu_R)' \; , \; \beta = (\beta_F,\beta_R)' \; , \;  \Omega = \mathrm{diag}(0,\Omega_R),
\label{eq:decomposition}
\end{equation}
where the notation $\beta = (\beta_F,\beta_R)'$ (resp. $\mu = (\mu_F,\mu_R)'$) means that the matrix $\beta$ (resp. $\mu$) is given by the concatenation of the two matrices $\beta_F$ and $\beta_R$ (resp. $\mu_F$ and $\mu_R$) one below the other, leading to :
\begin{equation*}
g
\begin{pmatrix}
\psi_{F,i}\\\psi_{R,i}
\end{pmatrix} = \begin{pmatrix}
\mu_{F}\\\mu_{R}
\end{pmatrix} +  \begin{pmatrix}
\beta_{F}\\\beta_{R}
\end{pmatrix} C_i + \begin{pmatrix}
0\\\eta_{R,i}
\end{pmatrix} \; , \; \eta_{R,i} \sim \mathcal{N}(0,\Omega_R).
\end{equation*}

\subsubsection{Appropriate BIC penalties}

Let $\ell(\mathbf{y};\hat{\mu},\hat{\beta},\hat{\Omega}_R)$ denote the log-likelihood of the observations computed at the maximum likelihood estimates of the model parameters. 

\begin{prop} 
Assume that $S_{\Omega}$ is given, then the appropriate BIC expression for covariate selection is given by:
\begin{equation}
BIC_{\beta} = -2\ell(\mathbf{y};\hat{\mu},\hat{\beta},\hat{\Omega}_R) + \dim(\{\mu_R,\beta_{R}\}) \log N+ \dim(\{\mu_F,\beta_{F}\}) \log n_{\mathrm{tot}},
\end{equation}
where $\dim(A)$ counts the number of non-zero elements of $A$. $\mu_R$, $\mu_F$, $\beta_R$ and $\beta_F$ are defined in \eqref{eq:decomposition}, $N$ is the number of subjects and $n_{\mathrm{tot}} = \sum_{i=1}^N n_i$ is the total number of observations.
\label{prop:BICcovariate}
\end{prop} 

\noindent A detailed proof of proposition \ref{prop:BICcovariate} can be found in \cite{Delattre2014}.

\begin{prop} 
Assume that $S_{\beta}$ is given, then the appropriate BIC expression for random effects selection is given by: 
\begin{equation}
BIC_{\Omega}=-2\ell(\mathbf{y};\hat{\mu},\hat{\beta},\hat{\Omega}_R)+\mathrm{card}(S_{\Omega})\log(N),
\end{equation}
where $\mathrm{card}(S_{\Omega})$ is the cardinal of $S_{\Omega}$.
\label{prop:BICcovariance}
\end{prop}

\begin{proof}
Proposition \ref{prop:BICcovariance} can be obtained by following very similar lines of proof as in \cite{Delattre2014}. Let us denote by $m$ any candidate parametric model for $\mathbf{y}$ with parameter $\theta_m=(\mu_m,\beta_m,\Omega_m)$. BIC derives from an asymptotic evaluation of $p(m|\mathbf{y})$ in a Bayesian framework. The key of this approximation is to evaluate the asymptotic behavior of the Laplace approximation of $p(\mathbf{y}|m)$. By standard intermediary calculations, one can show that this ultimately amounts to evaluate the asymptotic order of magnitude of $\log p(\mathbf{y}|\widehat{\theta_m},m) - \frac{1}{2} \log \det H_{\widehat{\theta_m}}$, where $\widehat{\theta_m}$ and $H_{\widehat{\theta_m}}$ respectively denote the maximum likelihood estimate of $\theta_m$ and the observed information matrix. Here, the only differences between the candidate models lie in their respective covariance structures defined above as $S_{\Omega}$. To be consistent with the asymptotic framework from which proposition \ref{prop:BICcovariance} is derived, we consider as in \cite{Delattre2014} the asymptotics defined by $N, n \rightarrow +\infty$, $n/N \rightarrow +\infty$. Under basic regularity assumptions on \eqref{eq:nonlinear} and \eqref{eq:phiGauss}, the Hessian $H_{\widehat{\theta_m}}$ behaves as $N$ times the information matrix of $m$, $I_{\theta}$, leading to $\frac{1}{2} \log \det (H_{\widehat{\theta_m}}) = \frac{\mathrm{dim}(\theta)}{2} \log(N) + \frac{1}{2} \log I_{\theta}$.  We conclude by using the asymptotic results from \cite{Nie2007} which state that when $\theta=\Omega$, the second term behaves as a constant.
\end{proof}

\noindent 
\textbf{Remarks} 
\begin{itemize}
\item Computing the maximum likelihood estimate and the log-likelihood are difficult problems in practice in many nonlinear mixed-effects models. Indeed, as the random effects $\eta_i$ are not observed, the expression of $\ell(\mathbf{y};\mu,\beta,\Omega_R)$ involves integrals over the distributions of the $\psi_i$ that often don't have any closed form expression:   
 $$
 \ell(\mathbf{y};\mu,\beta,\Omega_R)=\log \left[\prod_{i=1}^{N} \int{p(\mathbf{y}_i|\psi_{R,i},\mu_F,\beta_F)p(\psi_{R,i};\mu_R,\beta_R,\Omega_R) d\psi_{R,i}}\right].
 $$
Nevertheless, many methods have been developed to estimate the model parameters and to compute approximations of $\ell(\mathbf{y};\hat{\mu},\hat{\beta},\hat{\Omega}_R)$. They are implemented in mixed models softwares. 
\item Consistency of criteria $BIC_{\beta}$ and $BIC_{\Omega}$ remains to be proved. This is however not straightforward as classical proofs require the candidate models to be nested (\cite{Lebarbier2006}), which is not the case here.
\end{itemize}

\subsection{Stepwise selection algorithm}
\label{sec:stepwise}

%

Choosing the optimal model means selecting the optimal overall structure $(S_{\beta},S_{\Omega})$. The proposed algorithm alternates the selection of $S_{\Omega}$ and the selection of $S_{\beta}$ through inclusion and exclusion steps. At iteration $t$, an updated covariate structure $S_{\beta}^{(t)}$ and an updated covariance structure $S_{\Omega}^{(t)}$ are returned. 

\begin{center}
\begin{tabular}{p{13cm}}
\hline
\textbf{Algorithm.}\\
\hline
\\
\textbf{Initialization.} Let choose an initial model $M_0$ defined by covariate structure $S_{\beta}^{(0)}$ and covariance structure $S_{\Omega}^{(0)}$. Let $\beta^{(0)}$ and $\Omega_R^{(0)}$ be the fixed-effects parameter and the covariance matrix of $M_0$. \\
\\
\textbf{Iteration $t$.} Let $S_{\beta}^{(t)}$ and $S_{\Omega}^{(t)}$ respectively denote the covariate structure and the covariance structure from iteration $t-1$. 
\begin{enumerate}
\item \textbf{Selection of the random effects structure.} Consider all candidate covariance structures $(S_{\Omega}^k)_{k=1,\ldots,K}$ defined by the user. Each $S_{\Omega}^k$ corresponds to a covariance matrix $\Omega_{R,k}$. 
\begin{itemize}
\item Fit all $K$ models and compute the log-likelihoods $\ell(\mathbf{y},\widehat{\mu^{(t)}},\widehat{\beta^{(t)}},\widehat{\Omega}_{R,k})$, $k=1,\ldots,K$.
\item Choose the best among these $K$ structures and call it $S_{\Omega}^{(t+1)}$.
Here \textit{best} is defined as having smallest $BIC_{\Omega}$ from proposition 2.
\end{itemize}
\item \textbf{Selection of the covariate structure.} Consider the $L_t$ covariate structures obtained by adding or removing one covariate in $S_{\beta}^{(t)}$. Each covariate structure corresponds to a fixed-effects parameter $\beta_l$, $l=1,\ldots,L_t$.
\begin{itemize}
\item Fit all $L_t$ models and compute the log-likelihoods $\ell(\mathbf{y},\widehat{\mu_l},\widehat{\beta_l},\widehat{\Omega}_{R}^{(t+1)})$, $l=1,\ldots,L_t$.
\item Choose the best among these $L_t$ structures and call it $S_{\beta}^{(t+1)}$. 
Here \textit{best} is defined as having smallest $BIC_{\beta}$ from proposition 1. 
\end{itemize}
\item \textbf{Repeat 1. and 2.} until BIC values are no longer improved. 
\end{enumerate}\\
\hline
\end{tabular}
\end{center}

\medskip

\noindent 
\textbf{Remarks:}
\begin{itemize}
\item The collection of covariance structures remain the same over iterations. It is chosen by the user from its experience in the field of application. 
\item The overall procedure may be used in a purely forward or backward form for covariate selection. It may be initialized with non empty covariate structures. It is however recommended to start with simple covariate structure to avoid numerical difficulties that occur when inferring overly complex models. 
\end{itemize}

\section{Simulations}

We present numerical experiments to assess the performance and the robustness of the stepwise selection procedure of section \ref{sec:stepwise}. 

\subsection{Models}

We consider various scenarios corresponding to the following three models : a linear mixed effects model, a mixed effects Poisson model and a nonlinear mixed effects model. 

\paragraph{Example 1.} We consider a linear mixed effects model with $p$ covariates and $d=3$ individual parameters $\psi_i=(\psi_{i1},\psi_{i2},\psi_{i3})'$ :
\begin{eqnarray}
y_{ij} & = & \psi_{i1} x_{1ij} + \psi_{i2} x_{2ij} + \psi_{i3} x_{3ij} + \epsilon_{ij} \; , \; \epsilon_{ij} \underset{i.i.d.}{\sim} \mathcal{N}(0,\sigma^2), \nonumber\\
\psi_i & = & \mu + \beta C_i + \eta_i \; , \; \eta_i \underset{i.i.d.}{\sim} \mathcal{N}_3(0,\Omega),
\label{eq:simulin}
\end{eqnarray} 
where $i=1,\ldots,N$ stands for the individual, $j=1,\ldots,n$ denotes the index of the $j^{th}$ observation for a given individual, and $x_{1ij}$, $x_{2ij}$ and $x_{3ij}$ are regression variables that are part of the structural model. $\mu$ is the intercept, $C_i$ is the vector of $p$ covariates for individual $i$ and $\beta$ is a $d\times p$ matrix of fixed-effects. For the sake of simplicity, we considered a diagonal covariance matrix for the random effects $\Omega=\textrm{diag}(\omega_1^2,\omega_2^2,\omega_3^2)$. The sequences $(\eta_i)$ and $(\epsilon_{ij})$ are assumed to be mutually independent. 

\noindent Note that we can rewrite model \eqref{eq:simulin} in the usual formalism of linear mixed effects models :
\begin{equation*}
y_i = X_i \delta + Z_i \eta_i + \epsilon_i,
\end{equation*}
where $X_i = Z_i \otimes \tilde{C_i}$, $\tilde{C_i}=\begin{pmatrix} 1 & C_i'\end{pmatrix}$ and $\delta = (\mu_1,\beta_{11},\ldots,\beta_{1p},\ldots,\mu_d,\beta_{d1},\ldots,\beta_{dp})'$.  
Here, $Z_i =\begin{pmatrix}x_{1i1} & x_{2i1} & x_{3i1} \\
\vdots & \vdots & \vdots \\
x_{1in_i} & x_{2in_i} & x_{3in_i}\end{pmatrix}$.

\paragraph{Example 2.} In this example we consider the following Poisson mixed effects model
\begin{equation}
y_{ij} | \psi_i \sim \mathcal{P}(e^{\psi_{i1} + \psi_{i2} x_{ij}}),
\label{eq:simGLMM}
\end{equation}
where $\psi_i=(\psi_{i1},\psi_{i2})'$ is defined as $\psi_i=\mu + \beta C_i + \eta_i $ , $\eta_i \underset{i.i.d.}{\sim} \mathcal{N}_2(0,\Omega)$ with $\Omega = \mathrm{diag}(\omega_1^2,\omega_2^2)$, with the same notations as in Example 1. 

\paragraph{Example 3.} We consider the following nonlinear mixed effects model which is widely used in pharmacokinetics for describing the evolution of drug concentration over time : 
\begin{equation}
y_{ij}=\frac{d_i ka_{i}}{V_i ka_{i}-Cl_{i}}\left[e^{-\frac{Cl_{i}}{V_i} t_{ij}} - e^{-ka_{i} t_{ij}}\right] + \epsilon_{ij}
, \quad \epsilon_{ij} \underset{i.i.d.}{\sim} \mathcal{N}(0,\sigma^2).
\label{eq:modelPK}
\end{equation}
$y_{ij}$ represents the measure of drug concentration on individual $i$  at time $t_{ij}$, $d_i$ is the dose of drug administered to individual i, and $V_i$, $ka_i$ and $Cl_i$ respectively denote the volume of the central compartment, the drug's absorption rate constant and the drug's clearance of individual $i$. Here the vector of individual parameters is $\psi_i = (ka_i,Cl_i,V_i)'$ where $\log(\psi_i)=\mu + \beta C_i + \eta_i $ where $\log(\cdot)$ should be understood component by component, $\eta_i \underset{i.i.d.}{\sim} \mathcal{N}_3(0,\Omega)$ with $\Omega = \mathrm{diag}(\omega^2_{ka},\omega^2_{Cl},\omega^2_{V})$. The sequences $(\eta_i)$ and $(\epsilon_{ij})$ are mutually independent. We emphasize that, contrary to $C_i$, $d_i$ and $t_{ij}$ are part of the structural model, thus they are not subject to selection.  \\

\noindent In each example, the model selection issue is to identify the nonzero components of $\beta$ and the nonzero components of $\Omega$.


\subsection{Simulation design}

\paragraph{Example 1.} 

We generate data with different values of $N$ and $n$ : $(N=20,n=10)$, $(N=100,n=10)$, $(N=20, n=50)$ and $(N=100, n=50)$. For each value of $(N,n)$, 500 datasets are simulated. The covariates $c_i^{(1)},\cdots, c_i^{(p)}$ and the regression variables $x_{1ij}$, $x_{2ij}$ and $x_{3ij}$ are randomly drawn independently from a uniform distribution over the interval $[-4,4]$. We set the intercept parameters $\mu_1=\mu_2=\mu_3=0.5$.\\

\noindent We consider two scenarios with $p=2$ and $p=8$ covariates.
\begin{enumerate}
\item When $p=2$, the true fixed effects are $\beta = \begin{pmatrix} 1 & 0 & 0 \\ 0 & 1 & 0 \end{pmatrix}'$.
\item When $p=8$, $\psi_{i1}$ depends on covariates $c_i^{(1)}$ and $c_i^{(2)}$ ({\it ie} the first two elements of the fixed effects matrix are $\beta_{11}=\beta_{12}=1$ and the remaining elements of the first line are all zero), $\psi_{i2}$ depends on covariates $c_i^{(1)}$, $c_i^{(3)}$ and $c_i^{(4)}$ and $\psi_{i3}$ depends on covariate $c_i^{(5)}$ ($\beta_{21}=\beta_{23}=\beta_{24}=\beta_{35}=1$, the remaining $\beta$ elements are all zero). 
\end{enumerate}

\noindent For each scenario, two covariance settings are considered. 

\begin{enumerate}
\item \textit{Large variance.} The first one displays high variance parameters, that should favorize the identification of the true random effects on $\psi_{i1}$ and  $\psi_{i3}$ : $\omega_1^2=3$ and $\omega_3^2=2$. 
\item \textit{Small variance.} The second one displays one random effect with a very small variance parameter : $\omega_1^2=3$ and $\omega_3^2=0.05$. Thus, the random effect on $\psi_{i3}$ is less easily detectable.
\end{enumerate}

\paragraph{Example 2.}

As in Example 1, we use different values for $N$ and $n$ leading to different sample sizes: $(N=20,n=10)$, $(N=100,n=10)$, $(N=20, n=50)$ and $(N=100, n=50)$ and for each value of $(N,n)$, 500 datasets are simulated. Here, the covariates $c_i^{(1)}$ and $c_i^{(2)}$ and the regression variable $x_{ij}$ are independently randomly drawn from uniform distributions : $x_{ij} \underset{i.i.d.}{\sim}\mathcal{U}(-1.5,1.5)$, $c_i^{(1)} \underset{i.i.d.}{\sim} \mathcal{U}(-3,3)$ and $c_i^{(2)} \underset{i.i.d.}{\sim} \mathcal{U}(-2,2)$. For this experiment, we set $\beta_{11}=\beta_{22} = 0.5$ and the remaining elements of $\beta$ are all zero, $\mu_1=0.1$, $\mu_2=0$, $\omega_1^2=1$ and $\omega_2^2=0$.

\paragraph{Example 3.}

In this experiment, we only vary the number of subjects : $N=30$ and $N=80$. We consider $10$ observations per individual and the observation times are: $t_{ij}=t_j \in (0.25,0.5,1,2,3.5,5,7,9,12,15)$. We use the same dose $d_i=320$ for all the individuals. The true values of the model parameters are : $V=29$, $ka=0.10$, $Cl=2.78$ where $\mu = (\log(ka),\log(Cl),\log(V))'$, $\Omega = \mathrm{diag}(0.3,0.1,0)$, $\beta_{11}=0.12$, $\beta_{32}=0.05$ and the remaining elements of $\beta$ are all zero. The covariates $c_i^{(1)}$ and $c_i^{(2)}$ are randomly drawn according to independant uniform and Bernoulli distributions : $c_i^{(1)} \underset{i.i.d.}{\sim} =\mathcal{U}(6,8)$ (continuous covariate) and $c_i^{(2)} \underset{i.i.d.}{\sim} = \mathcal{B}(1/2)$ (discrete covariate).


\subsection{Algorithms in competition}

We compared our stepwise selection algorithm (denoted by SSA) with an exhaustive search of the best model and available algorithms in the literature. 

\begin{enumerate}
\item The exhaustive search identifies the best model by minimizing a classical version of BIC over all possible models. The classical BIC penalty is given by $D \times \log(N)$ where $D$ is the number of estimated parameters. Here, with the notations of propositions \ref{prop:BICcovariate} and \ref{prop:BICcovariance},  $D = \mathrm{dim}(\{\mu,\beta\})+\mathrm{card}(S_{\Omega})$. Note that $BIC_{\beta}$ and $BIC_{\Omega}$ defined in Propositions \ref{prop:BICcovariate} and \ref{prop:BICcovariance} are not suitable because they are designed for comparison of models with either the same random effects covariance structure $S_{\Omega}$ or the same fixed-effects structure $S_{\beta}$. 

In the simulation study, we use the exhaustive search as a gold standard for assessing performance. While the exhaustive search is a simple and conceptually appealing approach,  it suffers from computational limitations and is only tractable when the number of covariates $p$ is small. We didn't implement it in Example 1 with $p=8$ since millions of models should have been compared.
\item The regularized penalized quasi-likelihood (rPQL) approach of \cite{Hui2017} optimizes a criterion defined as the sum of three terms: a penalized quasi-likelihood, an adaptive lasso penalty for the fixed effects and an adaptive group lasso penalty for the random effects. This method is however only feasible in generalized linear mixed effects models and was compared to our algorithm in Examples 1 and 2.
\end{enumerate}


We also compared our SSA selection algorithm to the procedure of \cite{Bondell2010}, which embeds an adaptive LASSO penalty in an EM algorithm, but we do not report the results here. The computing times related to Bondell's procedure are considerably longer and are prohibitive when the sample size is high. When the two algorithms were run on the same simulated data, we noted that our SSA algorithm outperformed the adaptive LASSO procedure in terms of its ability to recover the true model structures. We suspect that it is because LASSO penalties are inherently designed for high dimension settings where the number of variables is bigger than the total sample size.


\subsection{Results}

The calculations were started in batch mode on a processor Intel 12 core 2.1 Ghz. The results are given in Tables \ref{tab:lin2} and \ref{tab:lin8} for Example 1, in Table \ref{tab:glm} for Example 2 and in Table \ref{tab:nlme} for Example 3. Each table compares our proposed algorithm (denoted by SSA) to the exhaustive BIC search and the rPQL algorithm. We report in columns C, R and CR, the proportion of times the true structures $S_{\beta}^{\star}$, $S_{\Omega}^{\star}$ and both structures simultaneously $(S_{\beta}^{\star},S_{\Omega}^{\star})$ are selected by each method, respectively. In column "time" we give the mean computation times for each algorithm. Each line corresponds to increasing sample sizes $(N,n)$. \\

In every experiment and in all examples, we see that the stepwise selection procedure provides very similar performances to the exhaustive BIC search while offering some substantial reductions in computation time. In Tables \ref{tab:lin2}, \ref{tab:glm} and \ref{tab:nlme} ($p=2$ covariates), the percentage of times the true model is selected (CR) is greater than 75\%  in Examples 1 and 2 and is more than 96\% when $N=100$. In Example 3, this percentage is equal to 59\% when $N=30$ and 86\% when $N=80$. In all cases, this percentage is highly correlated to the ability to recover the true covariate positions (C). The true random effects structure is correctly selected in more than 95\% of the time (R). As shown by the results obtained for Example 1 (Table \ref{tab:lin2}), the range of the variance parameters for the random effects does not affect the performances of our algorithm. Our stepwise selection algorithm was run several times with different initializations of the algorithm, including more or less covariates in the initial model $M_0$. We obtained similar percentages C, R and CR. Thus the SSA algorithm is not very sensitive to the initialization, which is advantageous when there is no a priori idea on the truly informative covariates and random effects. \\

In Table \ref{tab:lin8}, the number of covariates is increased. This allows to investigate the performance when the number of competing covariate structures is huge. In Table \ref{tab:lin2}, there are 64 possible covariate structures while in Table \ref{tab:lin8} there are over 16 millions possible covariate structures. In this high dimension setting, rPQL produces better results to the ones of SSA which is expected since it is based on a regularized criterion that shrinks small coefficients towards 0. In the \textit{small variance} setting, however, SSA performs better in large samples and recovers the true model about 75\% of the time. Let us notice that rPQL is designed to perform selection of both design variables ($x_{ij}$) and covariates. In our problem formulation, the design variables are not subject to selection. The percentages (columns C and CR) were computed on the $\beta$ covariate parameters only. In the Poisson mixed effects model of Example 2 with $p=2$ covariates (Table \ref{tab:glm}), SSA performs as well as the exhaustive search and better than rPQL. \\

The results for the nonlinear mixed effects model are given in Example \ref{tab:nlme}. We see that the performance of SSA algorithm is very close to the one of the exhaustive BIC search which is our performance objective. As the sample size increases, the percentage of correctly selected structures increases and recovers the true model 86\% of the time when $N=80$. We emphasize that the SSA algorithm is the first algorithm which implements joint selection in nonlinear mixed effects models.\\

From these results we see that the SSA algorithm, which relies on appropriate BIC formulations for joint covariate and random effects selection, competes equally with the exhaustive search and presents a computationally efficient alternative in applications where the number of covariates and the number of random effects increase. In generalized linear mixed effects models, SSA and rPQL display comparable performances. However, SSA is the only available algorithm for joint selection in nonlinear mixed effects models.  

\begin{table}[p]
\centering
\caption{\label{tab:lin2} Linear mixed effects model with $p=2$ covariates. Proportion of datasets with correctly selected covariates (C), random effects (R) and both (CR) with the SSA algorithm, an exhaustive BIC search and the regularized PQL procedure of \cite{Hui2017}. The mean computation time, in seconds, is also provided for each method, with standard deviations in brackets.}
\bigskip
\footnotesize{}
\begin{tabular}{p{1cm}p{1.5cm}|cccl|cccl}
& & \multicolumn{4}{c|}{\textit{Setting 1}} & \multicolumn{4}{c}{\textit{Setting 2}}\\
& & \multicolumn{4}{c|}{\textit{Large variance}} & \multicolumn{4}{c}{\textit{Small variance}}\\
$(N,n)$ & Method & C  & R  & CR & time & C  & R  & CR & time\\
\hline
$(20,10)$ &  Stepwise & 0.92 & 0.95 & 0.88 & 1.13 (0.09) &  0.96 & 0.90 & 0.87 & 1.16 (0.11) \\ 
& Exh. BIC & 0.94 & 0.95 & 0.90 & 73.33 (8.02) & 0.96 & 0.90 & 0.86 & 75.26 (9.23)\\
& rPQL & 0.96 & 1.00 & 0.96 & 1.45 (0.13) & 0.53 & 0.62 & 0.34 & 2.59 (1.76) \\
\hline
$(20,50)$ & Stepwise & 0.89 & 0.97 & 0.87 & 1.99 (0.21) & 0.95 & 0.96 & 0.91 & 2.04 (0.19)\\
& Exh. BIC & 0.92 & 0.97 & 0.89 & 97.78 (11.57) & 0.96 & 0.96 & 0.92 & 99.97 (11.37) \\
& rPQL & 0.96 & 1.00 & 0.96 & 2.42 (0.22) & 0.51 & 0.94 & 0.48 & 2.95 (0.42)\\
\hline
$(100,10)$ & Stepwise & 0.98 & 0.99 & 0.97 & 2.26 (0.19) & 0.99 & 0.99 & 0.98 & 2.16 (0.14)\\ 
& Exh. BIC & 0.98 & 0.97 & 0.94 & 104.10 (11.73) & 0.99 & 0.97 & 0.96 & 104.40 (12.55) \\
& rPQL & 1.00 & 1.00 & 1.00 & 30.00 (1.35) & 0.70 & 0.90 & 0.61 & 39.53 (4.39)\\
\hline
$(100,50)$ & Stepwise & 0.99 & 0.99 & 0.98 & 7.48 (0.58) & 0.99 & 0.98 & 0.97 & 7.50 (0.50)\\
& Exh. BIC & 0.98 & 0.97 & 0.95 & 217.67 (21.65) & 0.99 & 0.98 &0.96 & 228.51 (21.49)\\
& rPQL & 1.00 & 1.00 & 1.00 & 115.59 (5.73) & 0.90 & 1.00 & 0.90 & 156.06 (8.78)\\
\end{tabular}
\normalsize{}
\end{table}

\begin{table}[p]
\centering
\caption{\label{tab:lin8} Linear mixed effects model with $p=8$ covariates. Proportion of datasets with correctly selected covariates (C), random effects (R) and both (CR) with the SSA algorithm, an exhaustive BIC search and the regularized PQL procedure of \cite{Hui2017}. The mean computation time, in seconds, is also provided for each method, with standard deviations in brackets.}
\bigskip
\footnotesize{}
\begin{tabular}{p{1cm}c|cccl|cccl}
& & \multicolumn{4}{c|}{\textit{Setting 1}} & \multicolumn{4}{c}{\textit{Setting 2}}\\
& & \multicolumn{4}{c|}{\textit{Large variance}} & \multicolumn{4}{c}{\textit{Small variance}}\\
$(N,n)$ & Method & C  & R  & CR & time & C  & R  & CR & time\\
\hline
$(20,10)$ &  Stepwise & 0.18 & 0.52 & 0.16 & 3.72 (1.90) &  0.26 & 0.48 & 0.23 & 3.75 (1.90) \\ 
& rPQL & 0.49 & 1.00 & 0.49 & 1.38 (0.15) & 0.22 & 0.30 & 0.01 & 5.58 (5.26) \\ 
\hline
$(20,50)$ & Stepwise & 0.10 & 0.33 & 0.09 & 5.60 (3.41) & 0.16 & 0.32 & 0.15 & 5.86 (3.54)\\
& rPQL & 0.49& 1.00 & 0.49 & 2.49 (0.28) & 0.10 & 0.64 & 0.02 & 6.06 (5.80)\\
\hline
$(100,10)$ & Stepwise & 0.63 & 0.99 & 0.62 & 10.01 (0.71) & 0.76 & 0.99 & 0.75 & 10.46 (0.72)\\ 
& rPQL & 1.00 & 1.00 & 1.00 & 27.36 (1.15) & 0.41 & 0.60 & 0.18 & 64.48 (34.10)\\
\hline
$(100,50)$ & Stepwise & 0.63 & 0.99 & 0.61 & 33.12 (2.85) & 0.74 & 0.99 & 0.73 & 36.41 (3.04)\\
& rPQL & 1.00 & 1.00 & 1.00 & 112.22 (5.76) & 0.52 & 0.93 & 0.48 & 205.40 (27.03)\\
\end{tabular}
\normalsize{}
\end{table}

\begin{table}[p]
\centering
\caption{\label{tab:glm} Poisson mixed effects model. Proportion of datasets with correctly selected covariates (C), random effects (R) and both (CR) with the SSA algorithm, an exhaustive BIC search and the regularized PQL procedure of \cite{Hui2017}. The mean computation time, in seconds, is also provided for each method, with standard deviations in brackets.}
\bigskip
\begin{tabular}{c|c|cccc}
$(N,n)$ & Method & C  & R  & CR & time \\
\hline
$(20,10)$ & Stepwise & 0.76 & 0.99 & 0.75 & 7.30 (1.37)\\
& Exhaustive BIC & 0.78 & 0.99 & 0.78 & 20.02 (0.56)\\
& rPQL & 0.64 & 0.99 & 0.64 & 2.65 (0.49) \\
\hline
$(20,50)$ & Stepwise & 0.73 & 0.99 & 0.73 & 15.02 (2.97) \\
& Exhaustive BIC & 0.79 & 0.99 & 0.79 & 34.86 (2.48)\\
& rPQL & 0.72 & 1 & 0.72 & 4.52 (0.85)\\
\hline
$(100,10)$ & Stepwise & 0.96 & 0.99 & 0.96 & 16.75 (1.89)\\
& Exhaustive BIC & 0.93 & 0.96 & 0.93 & 36.08 (1.90)\\
& rPQL & 0.87 & 1 & 0.87 & 11.89 (1.37)\\
\hline
$(100,50)$ &  Stepwise & 0.96 & 1.00 & 0.96 & 148.76 (18.44)\\
& Exhaustive BIC & 0.94 & 1.00 & 0.94 & 150.21 (11.26)\\
& rPQL & 0.87 & 1 & 0.87 & 35.91 (3.83)\\
\end{tabular}
\end{table}

\begin{table}[p]
\centering
\caption{\label{tab:nlme} Nonlinear mixed effects model. Proportion of datasets with correctly selected covariates (C), random effects (R) and both (CR) with the SSA algorithm and an exhaustive BIC search. The mean computation time, in seconds, is also provided for each method, with standard deviations in brackets.
}
\bigskip
\begin{tabular}{c|c|cccc}
$N$ & Method & C & R & CR & time\\
\hline
$N=30$ & Stepwise & 0.60 & 0.98 & 0.59 & 265.67 (50.47)\\
& Exhaustive BIC & 0.58 & 0.98 & 0.57 & 3589.30 (1345.22)\\
$N=80$ & Stepwise & 0.87 & 0.99 & 0.86 & 970.22 (210.69)\\
& Exhaustive BIC & 0.84 & 0.99 & 0.84 & 5503.182 (1686.36)\\
\end{tabular}
\end{table}

\section{Choosing the optimal amikacin dose : a model selection issue}

We resume the clinical pharmacology study published in \cite{Burdet2015}. In this study, $60$ patients with ventilator-associated pneumonia (VAP) received a 20 mg/kg single infusion dose of Amikacin. The Amikacin blood concentration was then measured at different times. Twelve covariates were also measured for each patient during the experiment, including the age ($age$), the sex ($sex$), the total body weight ($w$), the $PaO_2/FiO_2$ ratio which characterizes respiratory distress syndrome ($P/F$), the 4-h creatinine clearance ($ClCr$). The main objective of this work was to best characterize the pharmacokinetic variability between patients suffering from VAP in order to better choose the dose to be administered to patients. As the determination of dose is mainly based on Monte-Carlo simulations, the better the model describes individual variability, the more robust is the choice of dose. In \cite{Burdet2015}, the model design is based on a procedure described in \cite{Lavielle2007} associating Wald tests and likelihood ratio tests for covariate selection and BIC comparison for the selection of the covariance structure of the random effects. This procedure is quite usually used in PK, but there actually does not exist any clear consensus on model design in nonlinear mixed effects models frameworks. We illustrate our procedure on this concrete example.\\

We use the following two-compartments model to describe the evolution of Amikacin concentration: \\

\begin{equation*}
\begin{array}{l}
f(D,t,t_D, T_{inf}, \psi)=  \\
\frac{D}{T_{inf}} \left[\frac{A}{a} (1-e^{-\alpha(t-t_D)}) + \frac{B}{b} (1-e^{-\beta(t-t_D)})  \right] \textrm{ if } t - t_D \leq T_{inf} \\
\frac{D}{T_{inf}}  \left[\frac{A}{a} (1-e^{-aT_{inf}})e^{-a(t-t_D- T_{inf})} + \frac{B}{b} (1-e^{-b T_{inf}}) e^{-b(t-t_D- T_{inf})}  \right] \textrm{ otherwise}  
\end{array}
\end{equation*}
where 
\begin{spreadlines}{12pt}
\begin{alignat*}{2}
A = \frac{1}{V_1} \frac{a-Q/V_2}{a-b}\; ; \; B = \frac{1}{V_1} \frac{b-Q/V_2}{b-a} \; ; \; a = \frac{Q Cl}{V_1 V_2 b} \; ;\\
b = \frac{1}{2} \left[ \frac{Q}{V_1} + \frac{Q}{V_2} + \frac{Cl}{V_1} - \sqrt{\left(\frac{Q}{V_1} + \frac{Q}{V_2} + \frac{Cl}{V_1} \right)^2 - 4 \frac{Q}{V_2} \frac{Cl}{V_1}}\right].
\end{alignat*}
\end{spreadlines}
Here $\psi=(Q,Cl,V_1,V_2)$, $Cl$ is the Amikacin clearance, $V_1$ is the central volume of distribution, $V_2$ is the peripheral volume of distribution, and $Q$ is the inter-compartmental clearance. 

The observations ($y_{ij}$, $1 \leq j \leq n_i$) of individual $i$ are then modeled as follows:
\begin{equation*}
y_{ij} = f_{ij} + \left(u+v \, f_{ij}\right) \epsilon_{ij},
\end{equation*}
where $f_{ij}=f(D_i,t_{ij},t_{D,i}, T_{inf,i}, \psi_i)$, $x_{ij}=(D_i,t_{ij},t_{D,i}, T_{inf,i})$ are the regression variables, $\psi_i$ are the PK (pharmacokinetic) individual parameters for patient $i$, and the residual errors $\epsilon_{ij}$ are iid standard Gaussian random variables. Due to positivity constraints, $\psi_i$ are log-normal random variables, {\it i.e.} $\log \psi_i$ is Gaussian, such that $\log \psi_i = \mu + \beta C_i + \eta_i$, where $\eta_i \underset{i.i.d}{\sim} \mathcal{N}(0,\Omega)$ and $C_i$ is the vector of covariates for individual $i$.

We use the stepwise selection procedure of section \ref{sec:stepwise}. To avoid numerical difficulties due to model complexity and to the high number of available covariates, we chose to start the SSA algorithm with the null covariate model. As the number of random effects structures for $\psi_i$ is high, we chose to restrict the stepwise model research to diagonal covariance matrices and then tried to add some correlations between the random effects of the retained model in a second step.  We excluded the censored Amikacin data. As a consequence, we didn't exactly use the same data as in \cite{Burdet2015}. In this study, $N=53$ and $n_{tot}=247$. In order to allow model comparison a posteriori, Burdet's model is re-estimated based on this new dataset. Our model selection procedure led to a model that includes only three random effects (parameter $Q$ is not random) and no correlation between the three random effects:
\begin{eqnarray}
\log CL_i & = & \mu_{CL} + \beta_{CL,c} \, ClCr_i + \beta_{CL,a} \, age_i + \eta_{CL,i}, \nonumber\\
\log V_{1,i} & = & \mu_{V1} + \beta_{V1,w} \, w_i + \beta_{V1,P} \, P/F_i   + \eta_{V1,i}, \nonumber\\
\log Q_{i} & = & \mu_{Q} + \beta_{Q,s} \, sex_i, \nonumber\\
\log V_{2,i} & = & \mu_{V2} + \beta_{V2,c} \, ClCr_i + \eta_{V2,i}.
\label{eq:model1}
\end{eqnarray}

\noindent The main difference between our final model (model1) and the one retained in \cite{Burdet2015} (model2) is that model1 includes more covariates and less random effects than model2. Generally speaking in a mixed-effects model, the random effects aim at describing the part of the inter-individual variability that cannot be explained by the observed covariates. Model1 is quite satisfactory in this sense. Indeed, the included covariates allow to limit the complexity of the random effects structure whereas in model2, the covariance matrix of the random effects is full. Table \ref{tab:covariateAmikacin} summarizes the covariate structures of model1 and model2. In this table, symbol $\star$ identifies the significant covariates in model1 (see equation \eqref{eq:model1} above) and symbol $\bullet$ in model2. For instance in model1, the absence of random effect on parameter $Q$ can be explained by the fact that covariate $sex$ is significant to explain the between-subjects variability of $Q$. Adding the other covariates in model1 led to zeroing the correlation coefficients.

\begin{table}
\begin{center}
\caption{\label{tab:covariateAmikacin} Amikacin data: comparison of the covariate structures of model1 ($\star$) and model2 ($\bullet$).}
\begin{tabular}{c|c|c|c|c|c}
& ClCr & age & sex & w & P/F\\
\hline
CL & $\star$ $\bullet$ & $\star$ & & & \\
\hline
$V_1$ & & & & $\star$ $\bullet$ & $\star$ $\bullet$\\
\hline
Q & & & $\star$ & & \\
\hline
$V_2$ & $\star$ & & & & \\
\end{tabular}
\end{center}
\end{table}

\begin{table}
\begin{center}
\caption{\label{tab:modelcomparison} Amikacin data: model comparison.}
\begin{tabular}{|l|l|}
\hline
Model & BIC\\
\hline
model1 & 1492.624\\
model2 & 1643.676\\
model3 & 1513.123\\
\hline
\end{tabular}
\end{center}
\end{table}

We can compare model1 and model2 with an intermediate model3 that combines the random effects structure of model1 and the three covariates of model2. As these three models differ from both their covariate and covariance structures, BIC comparison relies on a classical BIC penalty given by $D \times \log(N)$ where $D = \mathrm{dim}(\{\mu,\beta\})+\mathrm{card}(S_{\Omega})$. The BIC values for the three models are given in Table \ref{tab:modelcomparison}. Model1 has the smallest BIC value. Simplifying the random effects structure of model2 strongly reduces the BIC value. This shows that \cite{Burdet2015} identified the most influent covariates but not the most relevant random effects.

\section{Concluding remarks}

In this paper, we propose an algorithm for joint selection of covariates and random effects in mixed-effects models. It is based on a formalism that includes linear and generalized linear models as well as nonlinear models. It can be easily extended to more complex models such as mixed-effects hidden Markov models that rely on a similar hierarchical formalism (\cite{Delattre2012}). We used a BIC approach to provide an easy to use algorithm for effects selection in low dimensional mixed models. From the simulation study, we see that it is very competitive with the existing methods both in terms of selection performance and in terms of calculation time. As any BIC procedure, our algorithm is not designed for the high dimensional setting where regularized methods are required both for model fitting and effects selection. Here, we provide a new practical method to deal with model choice issues that practitioners face in applications such as pharmacokinetics studies.

\section*{Aknowledgement}

The application was based on the data from the Impact trial (PI O. Pajot, Sponsor APHP, Paris, France, NCT00950222). We thank C. Burdet, O. Pajot,  C. Couffignal, L. Armand-Lef\`evre, A. Foucrier, C. Laou\'enan, M. Wolff, L. Massias and F. Mentr\'e for providing us the data. We are also grateful to the INRAE MIGALE bioinformatics platform (\url{https://migale.inrae.fr}) for providing computational resources.

\bibliographystyle{DeGruyter}

\bibliography{biblio}

\end{document}